%% file: main.tex
\documentclass[aps,twocolumn,superscriptaddress,footinbib,nobalancelastpage,pra]{revtex4-2}
\usepackage[utf8]{inputenc} % usually not needed (loaded by default)
\usepackage[T1]{fontenc}
\usepackage{amsmath,amssymb}
\usepackage[colorlinks=true,citecolor=blue,breaklinks]{hyperref}
\usepackage{graphicx}
\usepackage{slashed}
\usepackage{babel}
\usepackage[toc,page]{appendix}
\usepackage[table,xcdraw,dvipsnames]{xcolor}
\usepackage{bbm}
\usepackage{soul}
\usepackage{array,booktabs}

\newcommand*{\tr}{\mathbb{Tr}}% trace

\newcommand{\kb}[2]{| #1\rangle\!\langle #1|}

\newcommand{\be}{\begin{equation}}
\newcommand{\ee}{\end{equation}}
\newcommand{\diag}{\text{diag}}

\usepackage{natbib}
\usepackage{dsfont} % mathds{1} for identity
\usepackage{esint}     % integrals 
\usepackage{bm}% bold math
\usepackage{qcircuit}
\usepackage{bbold}
\usepackage{hhline}
\usepackage{float}
\usepackage{braket}
\usepackage{mathtools} % dcases
\usepackage{tabularx}
\usepackage{amsthm}
\usepackage{tikz}
\usepackage[explicit]{titlesec}

\newcommand{\Ha}{$F^G_{\rho_0}$ }
\newcommand{\Hb}{$F^G_{\rho_1} F^G_{\rho_2} $ }
\newcommand{\Hc}{$F_{\rho_2}^G F_{\rho_1}^G$ }

\def\w {\omega}
\def\bw {\bar{\omega}}

\begin{document}

\title{Unveiling the non-Abelian statistics of $D(S_3)$ anyons via photonic simulation}

\author{Suraj Goel}
\thanks{These authors contributed equally}
\affiliation{Institute of Photonics and Quantum Sciences, Heriot-Watt University, Edinburgh EH14 4AS, UK}

\author{Matthew Reynolds}
\thanks{These authors contributed equally}
\affiliation{School of Physics and Astronomy, University of Leeds, Leeds LS2 9JT, UK}

\author{ Matthew Girling}
\affiliation{School of Physics and Astronomy, University of Leeds, Leeds LS2 9JT, UK}

\author{Will McCutcheon}
\affiliation{Institute of Photonics and Quantum Sciences, Heriot-Watt University, Edinburgh EH14 4AS, UK}

\author{Saroch Leedumrongwatthanakun}
\affiliation{Institute of Photonics and Quantum Sciences, Heriot-Watt University, Edinburgh EH14 4AS, UK}

\author{Vatshal Srivastav}
\affiliation{Institute of Photonics and Quantum Sciences, Heriot-Watt University, Edinburgh EH14 4AS, UK}

\author{David Jennings}
\affiliation{School of Physics and Astronomy, University of Leeds, Leeds LS2 9JT, UK}
\affiliation{Department of Physics, Imperial College London, London SW7 2AZ, UK}

\author{Mehul Malik}
\email{m.malik@hw.ac.uk}
\affiliation{Institute of Photonics and Quantum Sciences, Heriot-Watt University, Edinburgh EH14 4AS, UK}

\author{Jiannis K. Pachos}
\email{j.k.pachos@leeds.ac.uk}
\affiliation{School of Physics and Astronomy, University of Leeds, Leeds LS2 9JT, UK}

\begin{abstract}
Simulators can realise novel phenomena by separating them from the complexities of a full physical implementation. Here we put forward a scheme that can simulate the exotic statistics of $D(S_3)$ non-Abelian anyons with minimal resources. The qudit lattice representation of this planar code supports local encoding of $D(S_3)$ anyons. As a proof-of-principle demonstration we employ a photonic simulator to encode a single qutrit and manipulate it to perform the fusion and braiding properties of non-Abelian $D(S_3)$ anyons. The photonic technology allows us to perform the required non-unitary operations with much higher fidelity than what can be achieved with current quantum computers. Our approach can be directly generalised to larger systems or to different anyonic models, thus enabling advances in the exploration of quantum error correction and fundamental physics alike.

\end{abstract}

\maketitle
{\bf \em Introduction:--} 
The exotic statistics of non-Abelian anyons make them of interest in fundamental physics \cite{Leinas,PhysRevLett.49.957,ALEXANDERBAIS199263,PhysRevD.48.4821,Srivastav19}. 
In addition, their resilience to local perturbations has given rise to several schemes for topological quantum computing and other applications~\cite{Deng17,Gavin07,A.Yu.Kitaev_2001,Raussendorf_2007,RevModPhys.80.1083}. 
This behavior is key for fault-tolerant quantum computing, making non-Abelian anyons a potential solution to error problems that limit the scaling of quantum computers \cite{kitaev03,Pachos12,RevModPhys.80.1083}. In the last decade we witnessed an intense effort to identify signatures of non-Abelian anyons in various physical platforms, such as FQH liquids at $\nu =5/2$~\cite{PhysRevLett.94.166802}, $p+ip$ topological superconductors~\cite{PhysRevB.61.10267,PhysRevLett.86.268} or quantum wires~\cite{A.Yu.Kitaev_2001}.  Unfortunately, the complexity of these systems allows for alternative interpretation of the observed signatures~\cite{Yu2021}. The conclusive characteristic of non-Abelian anyons is their exchange statistics, which is currently too complex to realise in the laboratory.

At the same time, several investigations have focused on simulating non-Abelian anyons \cite{Wootton_2017,JinShi,doi10.1126sciadv.aat6533,PRXQuantum.2.030323,Andersen22,Stenger22}. These efforts aim to establish the necessary conditions for observing non-Abelian statistics and addressing technical challenges in scaling and accuracy. Often, such simulations suffer from key loopholes. For example, the simulation of Majorana fermions utilizes a non-local encoding of fermion-like anyonic states in many qubits, with the help of the Jordan-Wigner transformation. However, this non-local encoding lacks the desired topological stability against local errors inherent in anyonic systems.

Here we propose and implement a photonic simulation that demonstrates the core features of non-Abelian anyon statistics corresponding to the $D(S_3)$ planar code~\cite{kitaev03}. Planar codes are both quantum error-correcting codes and condensed matter systems that host anyonic excitations. Although they require many-body interactions, their local encoding on spins makes them attractive for quantum simulations. The simplest version of the planar code is the toric code that supports Abelian anyons. The toric code has been already simulated in the laboratory with Josephson junctions~\cite{Acharya2023} and photonic systems~\cite{PhysRevLett.102.030502,Pachos_2009}.

We show that a single qutrit is sufficient to encode the core manipulations of $D(S_3)$ non-Abelian anyons and demonstrate their non-Abelian fusion and braiding properties. The operations required to generate and manipulate anyons are in general non-unitary matrices that have a unitary action on the anyonic Hilbert space~\cite{Luo11,PhysRevLett.101.260501,Brennen_2009}. The implementation of non-unitary operations is typically experimentally challenging with current quantum computing architecture. To overcome this problem we adopt photonic technologies that can perform non-unitary operations accurately and with high fidelity. This simulation can be expanded in two directions with advancements in technology. First, it can be scaled to larger lattice systems, allowing for a broader range of anyonic operations. Second, Hamiltonian interactions can be added to provide active fault-tolerance in the topologically encoded quantum information.

\begin{figure}
\includegraphics[width=0.9\linewidth]{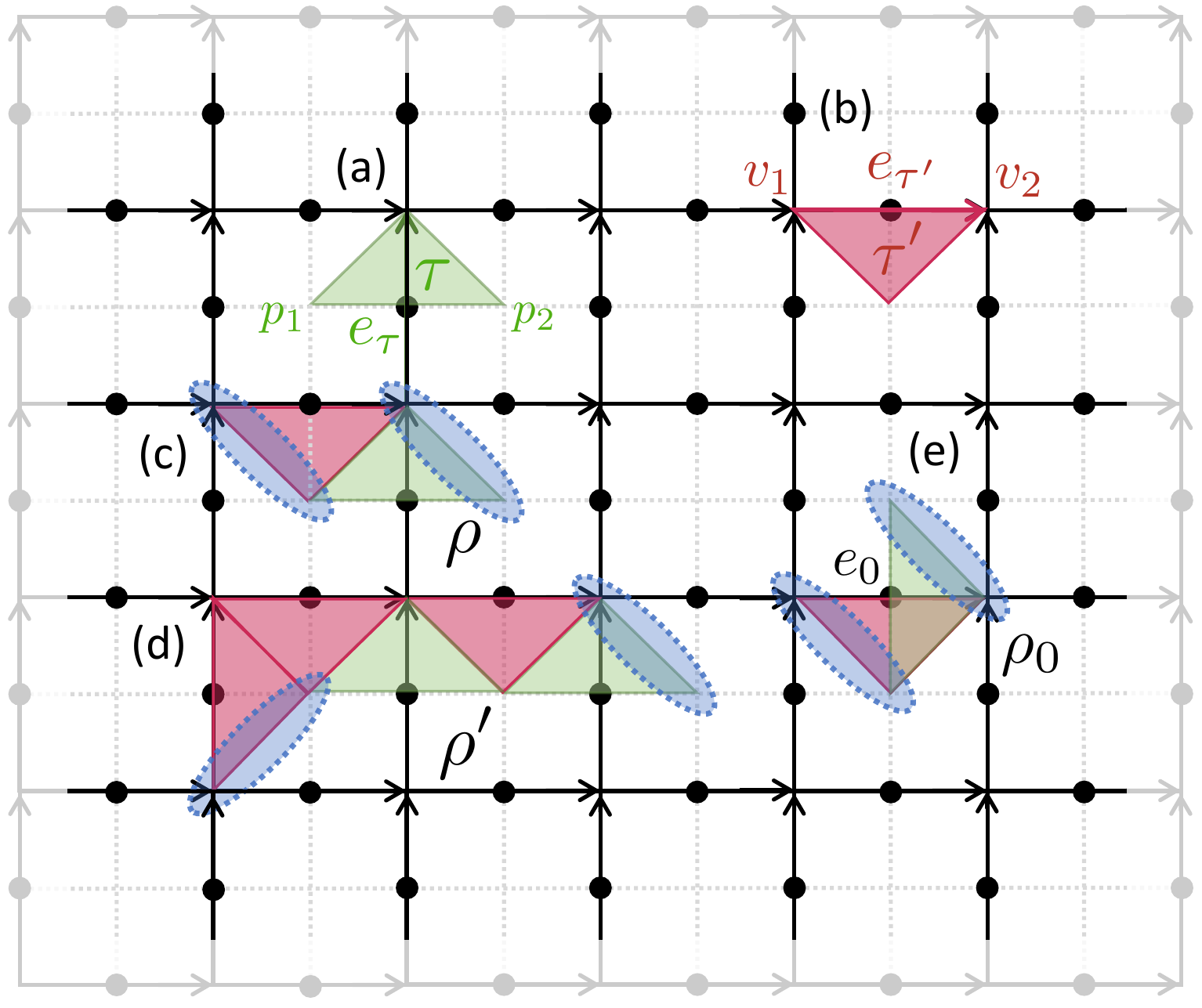}
\caption{Ribbon operators of $G$ anyons. (a) A dual triangle $\tau$ has support on link $e_\tau$ and creates excitations at $p_1$ and $p_2$ plaquettes. (b) A direct operator $\tau'$ has support on link $e_{\tau'}$ and create excitations at $v_1$ and $v_2$ vertices. (c,d) Compositions of dual and direct triangles give rise to ribbon operators $\rho$. Anyonic excitations of ribbon operators are dyons (dotted ovals) that are positioned at the endpoint plaquettes and vertices. (e) We define the elementary ribbon $\rho_0$ where both dual and direct triangles have support on the {\em same} link $e_0$.
}
\label{fig:lattice}
\end{figure}

{\bf \em The non-Abelian $D(S_3)$ anyonic model:--} The $D(S_3)$ model is based on the group transformations of a triangle, $S_3= \{e, c, c^2,t, tc, tc^2\}$, where $e$ is the identity element, $c$ the generator of $2\pi/3$ rotations and $t$ the generator of reflections. The $D(S_3)$ planar code consists of a square lattice where $d=6$ qudits are positioned at its links parameterised by the group elements of $S_3$, as shown in Fig. \ref{fig:lattice}. The Hamiltonian of the model has mutually commuting plaquette and vertex operators~\cite{kitaev03}. The ground state of the model is identified as the vacuum and its anyons are manifested as localised excitations at the vertices and/or the plaquettes. 

This $D(S_3)$ planar code supports eight different anyonic excitations labelled by $\{A, B, C, D, E, F, G, H\}$~\cite{Beigi2011,PhysRevB.96.195150,Wootton18}. Particle $A$ corresponds to the vacuum that fuses trivially with the rest of the anyons. Here we restrict ourselves to the $\{A, B, G\}$ subgroup that is closed under fusion, $B\times B =A$, $G\times B=G$ and $G\times G = A+B+G$. Moreover, $G$ has non-trivial braiding statistics. Its corresponding fusion, $F^G_{GGG}$, and braiding, $R^{GG}$, unitary matrices are given by
\begin{equation}
    F^G_{GGG} = \frac{1}{2}
\begin{pmatrix}
    1 & 1 & \sqrt{2} \\
    1 & 1 & -\sqrt{2} \\
    \sqrt{2} & -\sqrt{2} & 0
\end{pmatrix},\,\,
    R^{GG} = 
\begin{pmatrix}
    \bar\omega & 0 & 0 \\
    0 & \bar\omega & 0 \\
    0 & 0 & \omega
\end{pmatrix},
\label{grmatrix}
\end{equation}
where $\omega = e^{2\pi i/3}$,
which give a non-trivial braiding matrix $B^{GG} = FR^2F^\dagger$. The non-Abelian character of the $G$ anyons is manifested in the non-trivial commutation relation between $F^G_{GGG}$ and $(R^{GG})^2$. The $\{A, B, G\}$ subgroup should be contrasted to $\{A, B, C\}$, which has similar fusion rules but trivial braiding statistics, $B^{CC}=\mathbbm{1}$~\cite{PhysRevB.96.195150,PhysRevX.4.011051}.

It is possible to verify fusion and braiding properties of the planar code anyons by generating and manipulating the corresponding anyonic excitations. Such excitations are created from the vacuum state by applying operations on the links of the lattice. In general, these rotations are given in terms of ribbon operators $F_\rho^X$, where $\rho$ is the path along which the rotations are applied, giving rise to two $X$ anyons at its endpoints. These ribbon operators, together with their action on the ground state, encode the anyonic fusion and braiding properties. 

To define the ribbon operators we employ the oriented lattice representation of $D(S_3)$ shown in Fig. \ref{fig:lattice}. A dual triangle $\tau$ has support on a link $e_\tau$ and is connecting two plaquettes $p_1$ and $p_2$ adjacent to the link $e_\tau$, as shown in Fig. \ref{fig:lattice}(a). A direct triangle $\tau'$ has support on a link $e_{\tau'}$ and is connecting two vertices $v_1$ and $v_2$ adjacent to the link $e_{\tau'}$, as shown in Fig.~\ref{fig:lattice}(b). We now assign a six-dimensional Hilbert space $\{|h\rangle, h\in S_3\}$, to each link $e_\tau$. To every triangle $\tau$ we define an operator $L^h_{\tau_\text{dual}} = \sum_{g\in S_3}|hg\rangle\!\langle g|$, with $h\in S_3$, acting on $e_\tau$, if $e_\tau$ points towards $v$. Otherwise $L^h_{\tau_\text{dual}} = \sum_{g\in S_3}|gh^{-1}\rangle\!\langle g|$. Similarly, we define $P^g_{\tau_\text{dir}} =|g\rangle\!\langle g|$, with $g\in S_3$, if $e_\tau$ is clockwise w.r.t. $p$, otherwise $P^g_{\tau_\text{dir}} = |g^{-1}\rangle\!\langle g^{-1}|$. We next define the composite operators $F^{h,g}_\rho = L^h_{\tau_\text{dual}} P^g_{\tau_\text{dir}}$, where $\rho = \tau_\text{dir}\tau_\text{dual}$ with $\tau_\text{dir}$ a direct and $\tau_\text{dual}$ a dual triangle. Ribbon operators, $F^X_\rho$, that give rise to $X$ anyons are built out of such matrix elements, $F^{h,g}_\rho$, acting on qudits~\cite{PhysRevB.96.195150}. 

The $A$ and $B$ anyons are constructed from strings of operators corresponding to direct $\tau$'s. They give rise to anyons positioned at vertices, i.e. they have $h=e$ with $L^e=\mathbbm{1}$. Similar to the toric code's $e$ and $m$ anyons the $A$ and $B$ anyons can be created and moved around by applying single qudit unitary operator, $F^A_{\tau}$ and $F^B_{\tau}$, to a string of qudits~\cite{kitaev03}, where 
%$F^A_{\tau} = F^{e,e}_{\tau} + F^{e,c}_{\tau} + F^{e,c^2}_{\tau} + F^{e,t}_{\tau} + F^{e,tc}_{\tau} + F^{e,tc^2}_{\tau}$ and $F^B_{\tau} =  F^{e,e}_{\tau} + F^{e,c}_{\tau} + F^{e,c^2}_{\tau} - F^{e,t}_{\tau} - F^{e,tc}_{\tau} - F^{e,tc^2}_{\tau}$. The explicit representation of the $A$ and $B$ operators is given by \cite{kitaev03}
\begin{eqnarray}
F^A_{\tau} \!&=&\!\kb{e}{e}+\kb{c}{c} + \kb{c^2}{c^2} + \kb{t}{t} + \kb{tc}{tc} + \kb{tc^2}{tc^2},\,\,\,\,\nonumber\\
F^B_{\tau} \!&=&\! \kb{e}{e}+\kb{c}{c} + \kb{c^2}{c^2} - \kb{t}{t} - \kb{tc}{tc} - \kb{tc^2}{tc^2}.
\label{eq:FAFB}
\end{eqnarray}
Note that $F^A$ is the identity operator acting on the qudit and both $F^A$ and $F^B$ are unitary matrices whose sum, $F^A_{\tau} + F^B_{\tau} = 2(\kb{e}{e}+\kb{c}{c} + \kb{c^2}{c^2})$, can be described by a single qutrit operator.

The $G$ non-Abelian anyons are dyonic, i.e. it includes both direct and dual triangle operators forming ribbons. Dyons are positioned at composite plaquette and vertex at each of the endpoints of the ribbon, as shown in Fig. \ref{fig:lattice}(c,d,e). A simple way to create a pair of $G$ anyons on neighbouring vertices and plaquettes is to use the ribbon $\rho =\tau_\text{dual}\tau_\text{dir}$ \cite{Wootton18}, as shown in Fig.~\ref{fig:lattice}(c). The corresponding ribbon operator is given by
\be
F^G_\rho =  F^{c,e}_\rho + \omega F^{c,c}_\rho + \bar \omega F^{c,c^2}_\rho + F^{c^2,e}_\rho + \bar \omega F^{c^2,c}_\rho + \omega F^{c^2,c^2}_\rho,
\label{eq:small}
\ee
where for simplicity we omitted the overall normalisation. One can show that $F^G_\rho$ is hermitian but not unitary. Nevertheless, its restriction to the eigenstates of the $D(S_3)$ Hamiltonian are unitary \cite{Luo11}. Finally, note that larger ribbon operators, such as the ones shown in Fig. \ref{fig:lattice}(d), are created by acting simultaneously on all the relevant qudits with highly entangling operations~\cite{PhysRevLett.101.260501,Brennen_2009}.

{\bf \em Minimal encoding of non-Abelian anyons:--} To simplify the physical requirements for the simulation of $G$ anyons we identify the smallest possible ribbon that can encode the anyonic properties. It is possible to define a ribbon $\rho_0$, where the direct and dual triangles have support on the {\em same} qudit, as shown in Fig.  \ref{fig:lattice}(e). We define the corresponding ribbon operator as $F^G_{\rho_0} \equiv F^{c,e}_{\rho_0} + \omega F^{c,c}_{\rho_0} + \bar \omega F^{c,c^2}_{\rho_0} + \text{h.c.}$, where hermiticity is explicitly imposed. Explicitly, we have
\begin{eqnarray}
F^G_{\rho_0}=  |c\rangle\!\langle e| + \omega |c^2\rangle \! \langle c| + \bar \omega |e \rangle \! \langle c^2|  +\text{h.c.},
\label{eq:rib}
\end{eqnarray}
which acts only on three states.

With the minimal string and ribbon operators $F^A$, $F^B$ and $F^G$ we can explicitly verify the fusion properties of the $\{A, B, G\}$ subgroup of $D(S_3)$ by acting on a single qudit with six levels. By direct multiplication of the operators given in \eqref{eq:FAFB} and \eqref{eq:rib} we can verify their non-trivial fusion rules. In particular, when two ribbon operators, $F^G_{\rho_0}$, act on top of each other then the $G$ anyons at their endpoints are fused resulting to the ribbon operator of their fusion outcomes, i.e.
\be
F^G_{\rho_0} F^G_{\rho_0} =F^A_{\rho_0} + F^B_{\rho_0} + F^G_{\rho_0}.
\label{eq:rho1rho2}
\ee
This fusion process can be realised with a three level system as only the states $|e\rangle$, $|c\rangle$ and $|c^2\rangle$ are involved.

We next consider the braiding properties of $G$ anyons. In the case of the toric code the anyonic statistics of $e$ and $m$ anyons is given in terms of the commutation relations between their ribbon operators $F^e_{\rho_1} F^m_{\rho_2} = R^{em} F^m_{\rho_2} F^e_{\rho_1}$,
where $\rho_1$ and $\rho_2$ are two crossing paths of $e$ and $m$ anyons, respectively~\cite{kitaev03}. Due to topological invariance with respect to the exact shape of the path, the braiding relation can be realised by isolating the site, $\rho_0$, where paths $\rho_1$ and $\rho_2$ cross each other. As a result we can take the full system to be site $\rho_0$, with the anyons positioned outside the system's boundary. Then we have $F^e_{\rho_1\to \rho_0}=Z_{\rho_0}$ and $F^m_{\rho_2\to \rho_0}=X_{\rho_0}$ acting on the same qubit at $\rho_0$, thus obtaining the exchange statistics $R^{em}=-1$~\cite{PhysRevLett.102.030502,Pachos_2009}. 

For the $D(S_3)$ model the exchange of two $G$ ribbon operators takes the form~\cite{kitaev03}
\be
F^G_{\rho_1} F^G_{\rho_2} = R^{GG} F^G_{\rho_2} F^G_{\rho_1},
\label{eq:g}
\ee
where $R^{GG}$ is given by \eqref{grmatrix}. Hence, to determine $R^{GG}$ we need to implement $F^G_{\rho_1} F^G_{\rho_2}$ and $ F^G_{\rho_2} F^G_{\rho_1}$ and compare them. Similarly to the toric code case we employ a single site, $\rho_0$ and the minimal ribbon operator $F^G_{\rho_0}$, given in \eqref{eq:rib}, acting one it. As the operators we want to exchange are identical when acting on the single site system we adopt the following prescription. We first identify $F^G_{\rho_1} F^G_{\rho_2}$ with the product of two $F^G_{\rho_0}$ operators as given in \eqref{eq:rho1rho2}. Next, to determine $ F^G_{\rho_2} F^G_{\rho_1}$, we employ the exchange of their building blocks $ F^{h,g}_{\rho}$
\be
F^{h,g}_{\rho_2} F^{k,l}_{\rho_1} = F^{k,l\bar g \bar h g}_{\rho_1} F^{h,g}_{\rho_2},
\ee
valid for ribbons $\rho_1$ and $\rho_2$ with one common end. We employ these relations to compute $F^G_{\rho_2} F^G_{\rho_1}$ and then identify $\rho_1\to \rho_0$ and $\rho_2\to\rho_0$ to obtain (see Supplementary Material)
\begin{eqnarray}
F^G_{\rho_2} F^G_{\rho_1} 
= \bar \omega (F^A_{\rho_0} + F^B_{\rho_0}) + \omega F^G_{\rho_0}.
\label{eq:rho2rho1}
\end{eqnarray} 
Direct comparison of \eqref{eq:rho1rho2} and \eqref{eq:rho2rho1} deduces the desired braiding matrix $R^{GG}= \diag(\bar \omega, \bar \omega, \omega)$ in the $\{A,B,G\}$ basis.

Having a minimal system facilitates the simulation of braiding statistics with current technology. A natural first candidate is to employ current quantum computers to implement the ribbon operator, $F_{\rho_0}^G$. As $F_{\rho_0}^G$ is non-unitary it cannot be straightforwardly implemented with unitary quantum logic gates \cite{nielsen2002quantum}. However, as for any matrix, a unitary block encoding \cite{camps2022explicit,gilyen2019quantum,low2017optimal,low2019hamiltonian,camps2022fable} can be constructed where $F_{\rho_0}^G$ is embedded within a larger unitary, $U_F$. Through the preparation and measurement of a subset of qubits, a quantum circuit describing $U_F$ allows for $F_{\rho_0}^G$ to be applied to the labelled qutrit state. Based on the singular value decomposition of $F_{\rho_0}^G$ \cite{horn2012matrix}, $U_F$ must act on a minimum of 3 qubits (see Supplementary Material). Up to a rescaling of $F_{\rho_0}^G$, the success probability $p$ of implementing the transformation on pure states is bounded as $1/4 \le p \le 1$. Compiling an explicit $U_F$  using the native qiskit transpiler into the typical device gateset (e.g. single qubit rotations + CNOT) gives a circuit depth of 74 operations with 20 CNOTs \cite{Qiskit}. Using a simplified model of current device noise, we simulated the circuits applying this $U_F$ to states that maximise and minimise the success probability of applying $F_{\rho_0}^G$ to the labeled qutrit state (see Supplementary Material).  Unfortunately, current error rates are too high, and we observe low fidelities between the idealized circuits and the noisy implementations. As an alternative we resort to a photonic platform that can encode non-unitary operations in a straightforward way.

\begin{figure}[hbt!]
\centering\includegraphics[width=\columnwidth]{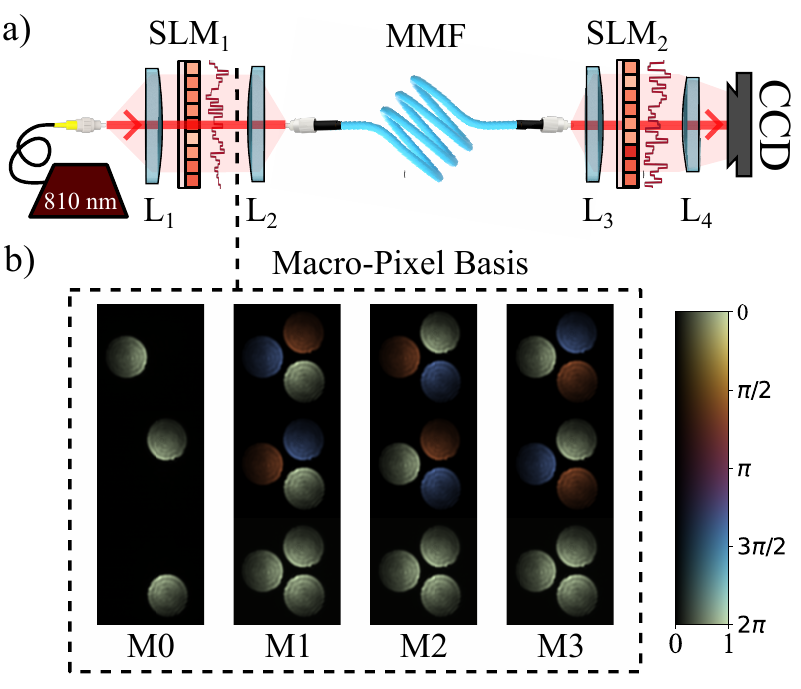}
\caption{(a) Experimental Setup. A coherent light source (810 nm) is incident on a phase-only spatial light modulator ($\text{SLM}_1$) and then coupled into a 2 m-long graded-index (GRIN) multi-mode fiber (MMF) with core diameter $50~\mu m$. 
The output of the MMF is incident on $\text{SLM}_2$ followed by a CCD camera. 
The combination of a high-dimensional mode mixer (MMF) sandwiched between two phase planes ($\text{SLM}_1$ and $\text{SLM}_2$) serves as a programmable optical circuit that can encode any non-unitary operations as shown in~\cite{goel2022inverse}. Additionally, $\text{SLM}_2$ is used for performing projective measurements required for quantum process tomography (QPT) to check the fidelity of the implemented transformations $\mathbf{T} \in \{F^G_{\rho_0}, F^G_{\rho_1}F^G_{\rho_2}, F^G_{\rho_2}F^G_{\rho_1}\}$. (b) Qutrit Encoding. Images showing three-dimensional photonic transverse-spatial modes in the macro-pixel basis (M0) generated by $\text{SLM}_1$. Modes from all mutually unbiased bases (M1, M2, M3) of the three-dimensional macro-pixel basis are also shown, which are used for performing QPT. 
}
\label{fig:apparatus}
\end{figure}

{\bf \em Experimental photonic simulation:--} Here we show how a photonic simulator can accurately perform the non-unitary operations needed to realise the fusion and braiding properties of the $G$ non-Abelian $D(S_3)$ anyons with mimimal errors. We experimentally implement the ribbon operator \Ha, that encodes two $G$ anyons at its endpoints, and its compositions \Hb and \Hc given in \eqref{eq:rib}, \eqref{eq:rho1rho2} and \eqref{eq:rho2rho1}, respecteively. The experiment consists of three distinct parts, as shown in Fig.~\ref{fig:apparatus}(a). First, we generate a qutrit state encoded in the transverse-spatial degree of freedom of light. We then evolve the state through the desired operations programmed on our photonic simulator. Finally, we characterise the implemented operations via quantum process tomography.
%Experimental figure

% Generation
The qutrit is encoded in a transverse-spatial modal basis consisting of discrete macro-pixels, as shown in Fig.~\ref{fig:apparatus}(b). We choose this particular basis as it can be tailored to perform high-quality projective measurements \cite{valencia2020high,SrivastavJTMA}. 
% Manipulation: Theory
Next, the qutrit state is evolved through the $G$ ribbon operator \eqref{eq:rib} and its compositions \eqref{eq:rho1rho2} and \eqref{eq:rho2rho1}, following which it is mapped onto spatially separated outcomes that can be measured on a camera. These operations are by definition non-unitary. Therefore, the task of performing an operation and sorting the outcomes spatially can be mapped to the problem of state discrimination between non-orthogonal states. 
%The correspondence comes in where each column of a given operator is mapped to a state vector. Performing a unitary operation corresponds to discriminating an orthogonal set of states, which is theoretically feasible with linear optics. 
%Similarly, the task of performing non-unitary operations can be mapped to discriminating a set of non-orthogonal states, but is theoretically infeasible to be performed perfectly with linear optics.
%Thankfully, t
There exist multiple schemes that perform this task by compromising either efficiency or accuracy of the discrimination~\cite{Barnett2009}. As we are interested in simulating the anyonic properties, we choose to enhance the accuracy of the operations at the expense of efficiency through optical losses. Herein we use the formalism of unambiguous state discrimination~\cite{Ivanovic1987, Chefles1998, CheflesBarnett1998, frankearnold12}, where one employs auxiliary modes to embed a low-dimensional non-unitary operation within a higher-dimensional unitary. The outcomes corresponding to the auxiliary modes can be ignored since they provide no information about the input state, and thus correspond to loss.

% It is well-known that it is impossible to perfectly implement non-unitary operations with linear optics.} To overcome this problem, we can either perform the operations efficiently, with inevitable loss in fidelity or accurately, at the expense of efficiency.
% Therefore, the problem of certifying the set of $\mathbf{T} \equiv \{$ \Ha, \Hb , \Hc$ \}$ operations can be mapped to high-dimensional unambiguous state discrimination \cite{goel2022simultaneously}. 
 %macro pixel modes? qdit encoding
% DESIGNING Operators

% MAnipulation, Experiment
Interestingly, the three operations we aim to implement, $\mathbf{T} \in \{F^G_{\rho_0}, F^G_{\rho_1}F^G_{\rho_2}, F^G_{\rho_2}F^G_{\rho_1}\}$, are Hermitian and have symmetric overlaps between different columns, i.e. $|T_{rj} \cdot {T_{r{i\neq j}}}^*| = \alpha, ~ \forall ~ r$  where $T_{rc}$ corresponds to the $r^{th}$ row and $c^{th}$ column of the given $\mathbf{T}$ matrix. This allows us to use a single auxiliary mode to perform these operations~\cite{Agnew2014} as was recently shown to be experimentally viable with optical circuits~\cite{goel2022simultaneously}. Due to the non-unitarity, these operations cannot be performed with unit success probability. Theoretically, the maximum average success probabilities are $50\%$, $37.5\%$ and $75\%$ for \Ha, \Hb and \Hc, respectively.    
% \AddA{This problem can be further simplified if the overlap between the optical modes is symmetric, i.e. $| \bra{\psi_i}\psi_{j\neq i}\rangle |^2=\alpha$. The three Hermitian operators $\mathbf{T} = \{$ \Ha, \Hb , \Hc$ \}$ are such that they have symmetric overlaps between different columns, i.e. $|T_{rj} \cdot {T_{r{i\neq j}}}^*|^2 = \alpha$ where $T_{rc}$ corresponds to the $r^{th}$ row and $c^{th}$ column of the given $\mathbf{T}$ matrix. This allows us to use a single auxiliary mode to perform these operations as recently shown in \cite{goel2022simultaneously}.
Using this approach, we encode our 3-dimensional Hermitian operators into 4-dimensional unitaries to proceed with this task.

These unitary operations are implemented using the recently demonstrated ``top-down'' approach, where an arbitrary optical circuit is embedded within a higher-dimensional mode-mixer sandwiched between two programmable phase planes \cite{goel2022inverse}. Our circuit uses a commercial multi-mode fibre (MMF) as a mode-mixer that is placed between two programmable spatial light modulators (SLMs) as shown in Fig.~\ref{fig:apparatus}(a). We use an inverse design technique known as the wavefront-matching (WFM) algorithm to program the SLMs. The WFM algorithm calculates the phase plane solutions by iteratively maximising the overlap between a set of input fields with the desired output ones. After updating the SLMs with the phase solutions given by the WFM algorithm, we couple a coherent light source with a wavelength of $810$ nm to characterize the implemented operation. The statistics of a single-photon qutrit state propagating through the system are identical to those obtained for a coherent state, allowing us to simplify the experiment and use a camera for detection~\cite{Barnett_2022}. 

\begin{table}[t!]
    \centering
    \begin{tabular}{| c || c || c |} 
        \hline
         ~~~Operation~~~ &  ~~~Fidelity~~~ &  ~~~Purity~~~  \\
         ($\mathbf{T}$) &   ($\mathcal{F}(\rho_{\mathcal{T}},\rho_{\widetilde{\mathcal{T}}})$) &   ($\mathcal{P}(\rho_{\widetilde{\mathcal{T}}})$) \\[1mm]
         \hline
 ~~\Ha~~ &~~$95.23 \pm 0.93 \%$~~&~~$96.04 \pm 0.03 \%$~~\\ [1mm] 
 ~~\Hb~~ &~~$94.44 \pm 0.85 \%$~~&~~$97.65 \pm 0.05 \%$~~\\ [1mm]
 ~~\Hc~~ &~~$97.59 \pm 0.59 \%$~~&~~$94.43 \pm 0.06 \%$~~\\ [1mm]
         \hline
    \end{tabular}
    \caption{Experimental results for the best case fidelity and purity of the processes corresponding to the \Ha, \Hb and \Hc operations. The error values are reported up to 3 standard deviations and correspond to systematic misalignment error (see Supplementary Material).}
     \label{tab:res}
\end{table}

We perform quantum process tomography to quantify the fidelity of the implemented operations $\mathbf{\widetilde{T}}$ in relation to the ideal operations $\mathbf{T}$.
Note that in addition to both SLMs being used for implementing the target operation $\mathbf{T}$, $\text{SLM}_1$ is used for generating the complete set of input modes (macropixel MUBs M0-M3, Fig.~\ref{fig:apparatus}(b)) and $\text{SLM}_2$ is used for performing the projective measurements needed for quantum process tomography (QPT).
% We use SLM$_1$ to prepare a set of input modes in all mutually unbiased bases (MUBs) in three dimensions as superposition of macro-pixel modes as shown in Fig.~\ref{fig:apparatus}(b) which evolve through the operation.
For each input mode, we measure the intensity at each of the three designated output modes at the camera, ignoring the auxiliary output. Next, SLM$_2$ is used to sequentially project the output into all mutually unbiased bases (MUBs) of the desired output modes. This is done in a manner similar to how projective measurements are performed with an SLM and a single mode fibre (SMF) \cite{Bouchard2018}, with the center region of the CCD camera used in place of the SMF. Using these measurements, we construct a coupling matrix between the complete set of input and output modes. This coupling matrix is then used to recover the implemented process via QPT, which we represent via its Choi state, $\rho_{\mathcal{\widetilde{T}}}= \mathcal{\widetilde{T}} \otimes \mathbb{1} (\rho^+)$, where $\rho^+$ is the maximally entangled state. The Choi state captures complete information about the process and can be used to evaluate the purity and fidelity to the target operations (See Supplementary Material) \cite{goel2022inverse}.

\begin{figure}[t!]
\centering\includegraphics[width=\columnwidth]{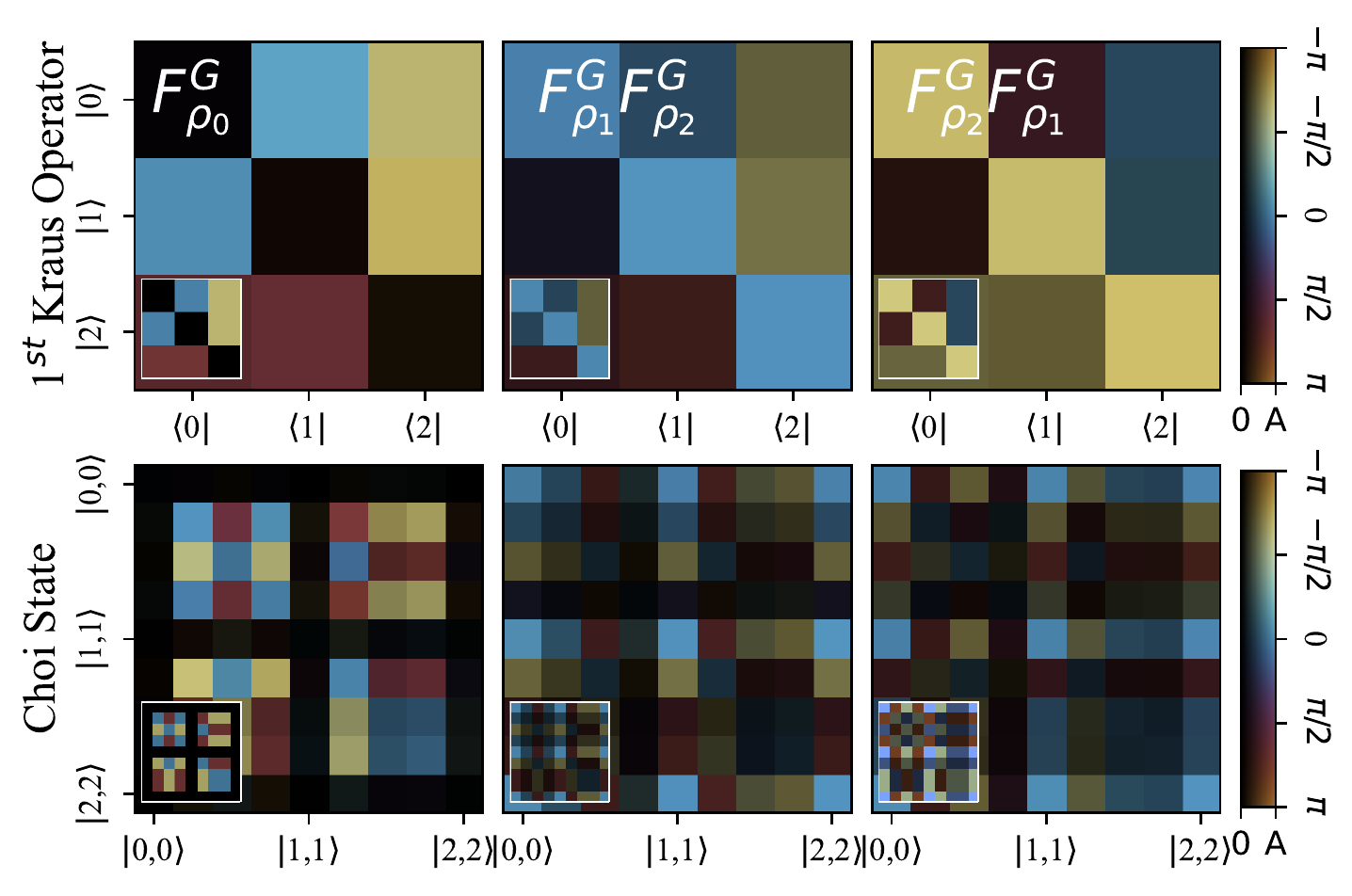}
\caption{Experimentally Measured Operators. We obtain the leading Kraus operator from the tomographed process (top row) and the Choi state representation for each operator \Ha, \Hb and \Hc (bottom row) (see Supplementary Material). Insets show theoretically expected results in each plot. $|A|$ corresponds to the maximum amplitude for a given plot and respective inset.}
\label{fig:results}
\end{figure}

We implement 10 realisations of each operation \Ha, \Hb and \Hc, and reconstruct their processes using the methods described above. Since different positions of measurement outcomes result in different performance, we vary the positions of measurement outcomes on the camera in each realisation in order to realize the best possible implementation of these operators. 
Out of all the implementations, the best-case fidelities, $\mathcal{F}$, and purities, $\mathcal{P}$, of the process for each operation \Ha, \Hb and \Hc  are shown in Table \ref{tab:res}. To visualise the quality of these operations, it is convenient to use the Kraus representation, which is an alternative way to represent these processes (see Supplementary Material). The ideal target operations have only one non-zero Kraus operator, which we can compare to the leading Kraus operators of the implemented processes owing to their high purity. Fig.~\ref{fig:results} (top row) depicts the leading Kraus operator of the implemented processes, showing that it agrees well with the target operations (insets). The full Choi state representation of these high purity processes is shown in Fig.~\ref{fig:results} (bottom row), with the ideal state shown for comparison.

% TO DO:  Challenges associated with separate operations, more complex operations/dims
%Explain QPT methods a bit more, MUBs (Fig 2)
%Description of figures, generation, characterisation of MMF (), measurements

{\bf \em Conclusions:--} Non-Abelian anyons present a fascinating and promising avenue for fault-tolerant quantum computation. Emulating their complex braiding statistics has so far evaded experimental realisation. In this letter, we have demonstrated a photonic simulation of the ribbon operators corresponding to $D(S_3)$ non-Abelian anyons with fidelities and purities above $94\%$. Our simulation has certified the minimal requirements and operations necessary to identify the statistics of these anyons, which can guide future efforts towards experimentally realising these exotic systems. Moreover, the extension of our experimental scheme to multi-qudit scalable quantum systems \cite{Erhard18,Paesani21,Cervera22} can potentially unlock the applications of non-Abelian anyons in quantum information processing. For example, two qutrits encoded in a nine-dimensional photonic system or in a scalable architecture with Josephson junctions or ion traps \cite{Ringbauer2022} could be employed to encode distinguishable $F^G_\rho$ ribbon operators where both the $F^G_{GGG}$ and $R^{GG}$ matrices can be realised. This work represents a significant step forward in the experimental study of non-Abelian anyons and opens the door to further exploration of their properties and potential applications.

\begin{acknowledgments}
We are grateful to Gavin Brennen, Sofyan Iblisdir and James Wootton for helpful discussions. This work was in part supported by EPSRC Grant No. EP/R020612/1.
\end{acknowledgments}

\bibliography{bibliographyAx}
\bibliographystyle{revtex4-1_longnames}

% \begin{thebibliography}{9}

% \bibitem{Kitaev03} 
% Kitaev, A. Y. {\em Fault-tolerant quantum computation by anyons.} Annals of Physics. {\bf 303}, 2-30 (2003).

% \bibitem{Wootton18}
% Laubscher, K., Loss, D. and Wootton, J.R. {\em Universal quantum computation in the surface code using non-Abelian islands.} Phys. Rev. A. {\bf 100}, 012338 (2019).

% \bibitem{Luo11}
%  X.-W. Luo, Y.-J. Han, G.-C. Guo, X. Zhou, and Z.-W. Zhou, Phys. Rev. A {\bf 84}, 052314 (2011).

% \end{thebibliography}
% \clearpage
% \appendix

% \newpage

\input{Appendix}

\end{document}

%% file: Appendix.tex
\clearpage
% \appendix

% \titlelabel{Appendix \thetitle}
% \titleformat{\section}{}{}{}{#1 \thesection}

\titleformat{\section}{\centering\bfseries}{#1}{1em}{\thesection}
\titleformat{\subsection}{\centering\bfseries}{ \thesection\thesubsection}{1em}{#1}

\renewcommand{\thesection}{\Alph{section}}
\renewcommand{\thesubsection}{.\arabic{subsection}}

\setcounter{equation}{0}
\setcounter{figure}{0}
\setcounter{table}{0}
\renewcommand{\theequation}{\thesection.\arabic{equation}}
\renewcommand{\thetable}{\thesection.\Roman{table}}
\renewcommand{\thefigure}{\thesection.\arabic{figure}}
\renewcommand{\theHfigure}{\thesection.\arabic{figure}}

\section{Appendix}
\label{append:A}

% \appendix
%\onecolumngrid

\begin{table*}[htp]
\begin{center}
\footnotesize
\begin{tabular}{ |c||c|c|c|c|c|c|c|c|c| } 
 \hline
 x & A & B & C & D & E & F & G & H \\ 
 \hline
 \hline
 A & A & B & C & D & E & F & G & H\\
 \hline
 B & B & A & C & E & D & F & G & H \\ 
 \hline
 C & C & C & A + B + C & D + E & D + E & G + H & F + H & F + G\\ 
 \hline
 D & D & E & D + E & A + C + F + G + H & B + C + F + G + H & D + E & D + E & D + E\\ 
 \hline
 E & E & D & D + E & B + C + F + G + H & A + C + F + G + H & D + E & D + E & D + E\\ 
 \hline
 F & F & F & G + H & D + E & D + E & A + B + F & H + C & G + C\\ 
 \hline
 G & G & G & F + H & D + E & D + E & H + C & A + B + G & F + C\\ 
 \hline
 H & H & H & F + G & D + E & D + E & G + C & F + C & A + B + H\\
 \hline
\end{tabular}
\end{center}
\caption{The fusion rules for $D(S_3)$, as calculated from the Verlinde formula.}
\label{table:fusions3}
\end{table*}

\begin{table*}[htp]
\begin{center}
\begin{tabular}{ |c|c||cccccc| } 
 \hline
 Anyon & Irrep & $e$ & $c$ & $c^2$ & $t$ & $tc$ & $tc^2$ \\ 
 \hline
 \hline
 1 & $\Gamma^{S_3}_1$ & 1 & 1 & 1 & 1 & 1 & 1 \\
 \hline
 $\Lambda$ & $\Gamma^{S_3}_{-1}$ & 1 & 1 & 1 & -1 & -1 & -1 \\ 
 \hline
 $\Phi$ & $\Gamma^{S_3}_{2}$ & $\begin{pmatrix} 1 & 0 \\ 0 & 1 \end{pmatrix}$ & $\begin{pmatrix} \bar\omega & 0 \\ 0 & \omega \end{pmatrix}$ & $\begin{pmatrix} \omega & 0 \\ 0 & \bar\omega \end{pmatrix}$ & $\begin{pmatrix} 0 & 1 \\ 1 & 0 \end{pmatrix}$ & $\begin{pmatrix} 0 & \omega \\ \bar\omega & 0 \end{pmatrix}$ & $\begin{pmatrix} 0 & \bar\omega \\ \omega & 0 \end{pmatrix}$ \\ 
 \hline

\end{tabular}
\end{center}

\caption{The irreducible representations of $\mathcal{N}_e = S_3$ \cite{PhysRevB.96.195150}.}
\label{table:abcirreps}
\end{table*}

%\twocolumngrid

\subsection{The $D(S_3)$ quantum double model}

We present here the non-Abelian group $S_3$, representing the transformations of a triangle and the $D(S_3)$ quantum double model. The $S_3$ group is given by 
\begin{equation}
     S_3 = \{e, c, c^2, t, tc, tc^2\}.
\end{equation}
where $c$ is the generator for rotations following $c^3 = e$, and $t$ the generator for reflections following $t^2 = e$ with $e$ the identity element. The non-Abelian nature of this group is defined by the relation $ct = tc^2$.

The quantum double for this group, $D(S_3)$, consists of eight anyons labelled $\{A,B,C,D,E,F,G,H\}$ \cite{Wootton18}, with $A$ corresponding to the vacuum. From the S matrix and Verlinde formula \cite{verlinde}, we can find the fusion rules of this group, given in Table~\ref{table:fusions3}. These fusion rules can be used to encode quantum information~\cite{jbook}. This information can be manipulated by anyonic braiding. Hence, quantum computation can be performed by anyonic fusion and braiding. Below we will investigate the braiding properties of particular anyonic subgroups of the $D(S_3)$ model.

\subsection{The \{1, $\Lambda$, $\Phi$\} Submodel}
The subgroup $\{A, B, C\}$ is also referred to as the $\{1, \Lambda, \Phi\}$ submodel \cite{jbook}. This consists of the vacuum and the two charge anyons of $D(S_3)$ corresponding to the irreducible representations of the trivial normaliser, $\mathcal{N}_e = S_3$: $\Gamma^{S_3}_1$, $\Gamma^{S_3}_{-1}$ and $\Gamma^{S_3}_{2}$ for $1$, $\Lambda$ and $\Phi$ respectively~\cite{PhysRevB.96.195150}. Here it is important to note that $\omega = e^{\frac{2\pi i}{3}}$ and $\bar\omega$ denotes its conjugate.

From Table~\ref{table:abcirreps} we obtain the fusion rules for this model
\begin{equation}
    1 \times 1 = \Lambda \times \Lambda = 1, \quad\quad \Phi \times \Lambda = \Phi, \quad\quad \Phi \times \Phi = 1 + \Lambda + \Phi.
\end{equation}
Anyon $\Phi$ is the only non-Abelian one with three fusion channels. It has been shown how to encode information in these anyonic fusion channels and how to manipulate it using non-topological operations.

To realise the fusion and braiding rules of the $D(S_3)$ quantum double model we first define ribbon operators, $F^1$, $F^\Lambda$ and $F^\Phi$, that give rise to the corresponding anyons at their endpoints. We can work out the form of the ribbon operators $F^X_\rho$ of abelian anyons $X$ at the endpoints of a path $\rho$ in the following way~\cite{qdabelian}
\begin{equation}
    F^{\chi,c}_\rho = \sum_{g\in S_3}\bar\chi(g)F^{\bar c,g}_\rho,
    \label{abelianbasistransform}
\end{equation}
where $c$ is an element of $S_3$ and $\chi$ is a character representation of $S_3$. In this case, the $c$ in equation (\ref{abelianbasistransform}) follows $c = e$ as $1$ and $\Lambda$ are of conjugacy class $C_e$. The characters are $1$ and $-1$ respectively. Therefore we need all ribbon operators of the group which have $h = e$, so all possibilities of $F^{e,g}_{\rho}, \ g\in S_3$. This gives the ribbon operators for $1$ and $\Lambda$ as
\begin{equation}
    F^1_\rho = F^{e,e}_{\rho} + F^{e,c}_{\rho} + F^{e,c^2}_{\rho} + F^{e,t}_{\rho} + F^{e,tc}_{\rho} + F^{e,tc^2}_{\rho}
\end{equation}
and
\begin{equation}
    F^\Lambda_\rho = F^{e,e}_{\rho} + F^{e,c}_{\rho} + F^{e,c^2}_{\rho} - F^{e,t}_{\rho} - F^{e,tc}_{\rho} - F^{e,tc^2}_{\rho}.
\end{equation}
These describe the ribbon operators $F^X$ creating $X$ anyons in terms of group ribbon operators $F^{h,g}$, which are defined as \cite{Wootton18,bombin2008family}
\begin{equation}
    F^{h,g}_\rho = L^h_{\tau_{dual}}T^g_{\tau_{dir}},
    \label{ribbonoperatordef}
\end{equation}
with
\begin{equation}
    L^g := 
    \begin{cases}
        \sum_{h \in S_3}\ket{gh}\bra{h} & \mathrm{if \ facing \ along \ edge.} \\
        \sum_{h \in S_3}\ket{hg^{-1}}\bra{h} & \mathrm{if \ facing \ away.}
    \end{cases}
    \label{loperators}
\end{equation}
\begin{equation}
    T^h := 
    \begin{cases}
        \ket{h}\bra{h} & \mathrm{if \ facing \ along \ edge.} \\
        \ket{h}\bra{h^{-1}} & \mathrm{if \ facing \ away .}
    \end{cases}
    \label{toperators} 
\end{equation}
For our derivations, we can assume the ribbons are all facing along an oriented edge. Using these definitions, we can rewrite the group ribbon operators as
\begin{equation}
    F^{e,e}_{\rho} = \ket{e}\bra{e}, \quad\quad F^{e,c}_{\rho} = \ket{c}\bra{c}, \quad\quad F^{e,c^2}_{\rho} = \ket{c^2}\bra{c^2},
\end{equation}
\begin{equation}
    F^{e,t}_{\rho} = \ket{t}\bra{t}, \quad\quad F^{e,tc}_{\rho} = \ket{tc}\bra{tc}, \quad\quad F^{e,tc^2}_{\rho} = \ket{tc^2}\bra{tc^2},
    \label{1lambdaphiribbondefs}
\end{equation}
which gives
\begin{equation}
    F^1 = \ket{e}\bra{e} + \ket{c}\bra{c} + \ket{c^2}\bra{c^2} + \ket{t}\bra{t} + \ket{tc}\bra{tc} + \ket{tc^2}\bra{tc^2} = \mathbb{1}
    \label{ro1}
\end{equation}
\begin{equation}
    F^\Lambda = \ket{e}\bra{e} + \ket{c}\bra{c} + \ket{c^2}\bra{c^2} - \ket{t}\bra{t} - \ket{tc}\bra{tc} - \ket{tc^2}\bra{tc^2}.
    \label{rolambda}
\end{equation}
To find the group ribbon operators corresponding to $F^\Phi_\rho$, we must introduce the non-Abelian basis transformation~\cite{bombin2008family}
\begin{equation}
    F^{RC;\mathbf{uv}} := \frac{n_R}{|\mathcal N_C|}\sum_{n\in \mathcal N_C} \bar\Gamma^{j,j'}_R(n)F^{\bar c_i,q_in\bar q_{I'}},
    \label{nonabelianbasistransform}
\end{equation}
with each anyon having a unique $C$, the conjugacy class and $R$, the irrep of the normaliser corresponding to $C$, $\mathcal N_C$. The unitary matrices of the irrep are given as $\Gamma_R(g), g\in S_3$ and the dimension of this is given by $n_R$. Each element of the normaliser and conjugacy class are given by $n_i$ and $c_i$ respectively. The representation $Q_C$ is defined as by $S_3/N_C$ and each element is given by $q_i$. Then $\mathbf{u} = (i,j)$ and $\mathbf{v} = (i',j')$ are extra degrees of freedom that can be traced out to give the final anyonic ribbon operator \cite{Wootton18}. Using this basis transformation, we find
\begin{equation}
    F^{\Phi}_\rho = 2F^{e,e}_\rho - F^{e,c}_\rho - F^{e, c^2}_\rho
\end{equation}
hence only three group ribbon operators are needed to define the $\Phi$ anyon. These were defined in (\ref{1lambdaphiribbondefs}) and can be substituted in to find
\begin{equation}
    F^\Phi_\rho = 2\ket{e}\bra{e} - \ket{c}\bra{c} - \ket{c^2}\bra{c^2}.
    \label{phiribbon}
\end{equation}

Now that the ribbon operators are defined, the braiding evolution can be found. For the Abelian anyons $1,\Lambda$ the braiding and fusion operations follow the familiar toric code steps. For the non-Abelian $\Phi$ anyons, it is simpler to braid ribbon operators in spacetime. For that we consider two loops $\rho_1$ and $\rho_2$ spanned by line paths traversed forwards and backwards so $\rho_1$ and $\rho_2$ braid with each other. This has the form~\cite{preskill}
\begin{equation}
    F^x_{\rho_1}F^x_{\rho_2}(F^x_{\rho_1})^{-1}(F^x_{\rho_2})^{-1}\ket{\psi} = U\ket{\psi},
\end{equation}
with $U$ being a unitary. To bring this state back to the vacuum and find $U$, we must know the commutation relation between $F^x_{\rho_1}$ and $F^x_{\rho_2}$ so we can cancel the ribbon operators applied to the same ribbon. For our $\Phi$ anyons, this means finding the result of swapping each group ribbon operator in (\ref{phiribbon}) using the exchange relation, describing the swapping of two anyons as having action~\cite{preskill}
\begin{equation}
    R : \ket{a,b} \longrightarrow \ket{aba^{-1},a},
    \label{qdrdef}
\end{equation}
which corresponds to the ribbon operator relation~\cite{qui}
\begin{equation}
    F^{h,g}_{\rho_1}F^{k,l}_{\rho_2} = F^{k,l\bar g\bar h g}_{\rho_2}F^{h,g}_{\rho_1}.
    \label{qdrribbondef}
\end{equation}
For the group ribbon operators of $\Phi$, this exchange gives
\begin{equation}
    F^{e,g}_{\rho_1}F^{e,l}_{\rho_2} = F^{e,l}_{\rho_2}F^{e,g}_{\rho_1} \quad \forall g, l \in S_3,
\end{equation}
so all of these commute. This means no phase is gained upon the double exchange of these anyons, and hence this subgroup has trivial braiding. 
This result could be also derived by the equation for the braiding matrix~\cite{Wootton18}
\begin{equation}
    B_{ab} = {F^d_{acb}}^{-1}R^2_{ab}F^d_{acb},
    \label{braidingfr2f}
\end{equation}
using the $F^\Phi_{\Phi\Phi\Phi}$ and $R_{\Phi\Phi}$ matrices given by~\cite{oldwootten}
\begin{equation}
F^\Phi_{\Phi\Phi\Phi} = \frac{1}{2}
    \begin{pmatrix}
        1 & 1 & -\sqrt{2} \\
    1 & 1 & \sqrt{2} \\
    -\sqrt{2} & \sqrt{2} & 0
    \end{pmatrix},
\quad
R_{\Phi\Phi} = 
    \begin{pmatrix}
        1 & 0 & 0 \\
   0 & -1 & 0 \\
   0 & 0 & 1
    \end{pmatrix},
    \label{en:fusion}
\end{equation}
in the basis $\ket{1}$, $\ket{\Lambda}$, $\ket{\Phi}$, giving
\begin{equation}
         B_{\Phi\Phi} = FR^2F^{-1} = 
\begin{pmatrix}
   1 & 0 & 0 \\
   0 & 1 & 0 \\
   0 & 0 & 1
\end{pmatrix} = \mathbb{1}_3.
\end{equation}
meaning there is no non-trivial evolution of the anyonic Hilbert space under braiding.

\subsection{The \{A, B, G\} Submodel}

A more promising submodel is given by the subgroup $\{A,B,G\}$. In this, $A$ and $B$ are the same $1$ and $\Lambda$ as before, but $G$ is a dyon. Therefore it is not an irrep of $S_3$ itself but of its normaliser $\mathcal{N}_c = \{e, c, c^2\} \cong \mathbb{Z_3}$ instead \cite{Wootton18}. The $G$ dyon corresponds to irrep $\Gamma^{\mathbb{Z}_3}_\omega$ shown in Table~\ref{table:abgirreps}.
\begin{table}
\begin{center}
\begin{tabular}{ |c|c||cccccc| } 
 \hline
 Anyon & Irrep & $e$ & $c$ & $c^2$ & $t$ & $tc$ & $tc^2$ \\ 
 \hline
 \hline
 $A$ & $\Gamma^{S_3}_1$ & 1 & 1 & 1 & 1 & 1 & 1 \\
 \hline
 $B$ & $\Gamma^{S_3}_{-1}$ & 1 & 1 & 1 & -1 & -1 & -1 \\ 
 \hline
 $G$ & $\Gamma^{\mathbb Z_3}_\omega$ & 1 & $\omega$ & $\bar\omega$ & 0 & 0 & 0 \\
 \hline
\end{tabular}
\end{center}

\caption{The irreducible representations of $\mathcal{N}_c = \mathbb Z_3$ \cite{PhysRevB.96.195150}.}
\label{table:abgirreps}
\end{table}
\par
Using Table~\ref{table:abgirreps} we can obtain the fusion rules for this model
\begin{equation}
    A \times A = B \cdot B = 1, \quad\quad G \times B = G, \quad\quad G \times G = A + B + G,
\end{equation}
with $G$ once again the only non-Abelian anyon. This model has identical fusion rules to the $\{1,\Lambda,\Phi\}$ submodel.

We now determine the braiding evolution of the $G$ anyons. We will first use equation (\ref{braidingfr2f}), for the $F^G_{GGG}$ and $R_{GG}$ matrices in the basis $\ket{A},$ $\ket{B},$ $\ket{G}$,
\begin{equation}
    F^G_{GGG} = \frac{1}{2}
\begin{pmatrix}
    1 & 1 & \sqrt{2} \\
    1 & 1 & -\sqrt{2} \\
    \sqrt{2} & -\sqrt{2} & 0
\end{pmatrix}
    \quad\quad\quad
    R_{GG} = 
\begin{pmatrix}
    \bar\omega & 0 & 0 \\
    0 & \bar\omega & 0 \\
    0 & 0 & \omega
\end{pmatrix}.
\label{grmatrix2}
\end{equation}
Substituting these into the braiding matrix equation (\ref{braidingfr2f}) gives
\begin{equation}
B_{GG} =
\begin{pmatrix}
    \cos{\frac{2\pi}{3}} & i\sin{\frac{2\pi}{3}} & 0 \\
    i\sin{\frac{2\pi}{3}} & \cos{\frac{2\pi}{3}} & 0 \\
    0 & 0 & e^{2\pi i/3}
\end{pmatrix},
\end{equation}
demonstrating that these dyons exhibit non-trivial non-Abelian braiding. 

The $A$ and $B$ ribbon operators have already been derived in equations (\ref{ro1}) and (\ref{rolambda}), with $F^A_\rho = F^1_\rho$ and $F^B_\rho = F^\Lambda_\rho$. For the $G$ anyon however, we need to use (\ref{nonabelianbasistransform}) to know which group ribbon operators to derive. Here we note that the conjugacy class of this dyon is $C_c = \{c, c^2\}$ \cite{PhysRevB.96.195150}. Performing this calculation gives the ribbon operator of $G$ as
\begin{equation}
F^G_\rho = F^{c,e}_\rho + \omega F^{c,c}_\rho + \bar\omega F^{c,c^2}_\rho + F^{c^2,e}_\rho + \bar\omega F^{c^2,c}_\rho + \omega F^{c^2,c^2}_\rho.
    \label{gribbon}
\end{equation}
For simplicity we omitted in this definition a normalisation factor of $1/3$ present in \eqref{nonabelianbasistransform}. From the general ribbon operator equation $F^{h,g}_\rho = L^h_{\tau_{dual}}T^g_{\tau_{dir}}$ we see that $F^G$ needs a dual and a direct triangle to be defined. In general, these are taken to be defined on different links of the lattice, such that the commutation relation 
\begin{equation}
L^h_{\tau_{dual}}T^g_{\tau_{dir}} =T^g_{\tau_{dir}}L^h_{\tau_{dual}}
\label{eq:comm}
\end{equation}
trivially holds as the $T$ and $L$ operators act on different Hilbert spaces. Due to that the Hermitian conjugate of $F^{h,g}_\rho$ is given by
\begin{equation}
\begin{split}
    (F^{h,g}_\rho)^\dagger &= (L^h_{\tau_{dual}}T^g_{\tau_{dir}} )^\dagger = (T^g_{\tau_{dir}} 
)^\dagger(L^h_{\tau_{dual}})^\dagger \\
& = T^g_{\tau_{dir}} L^{h^{-1}}_{\tau_{dual}}=
L^{h^{-1}}_{\tau_{dual}}T^g_{\tau_{dir}}=F^{h^{-1},g}_\rho
\end{split}
\label{eq:con}
\end{equation}
Here we want to identify the smallest possible system that can encode the fusion and braiding properties of the $\{A, B, G\}$ anyons. It is possible to define the dual and direct triangles to act on the {\em same} Hilbert space, i.e. we choose them to overlap (see main text). In this case the commutation property \eqref{eq:comm} does not hold any more and the conjugation relation given in \eqref{eq:con} is not satisfied automatically. To impose hermiticity of $F^G_\rho $ we take into account that $F^{c^2,e} = F^{c^{-1},e}= (F^{c,e})^\dagger$, $F^{c^2,c} = F^{c^{-1},c} = (F^{c,c})^\dagger$, $F^{c^2,c^2} = F^{c^{-1},c^2}= (F^{c,c^2})^\dagger$, so we can rewrite \eqref{gribbon} as
\begin{equation}
    F^G_\rho = F^{c,e} + \omega F^{c,c} + \bar\omega F^{c,c^2} + \text{h.c.}.
    \label{gribbon}
\end{equation}
Employing \eqref{ribbonoperatordef} we have
\begin{equation}
F^{c,e}_{\rho} = \ket{c}\bra{e}, \quad\quad F^{c,c}_{\rho} = \ket{c^2}\bra{c}, \quad\quad F^{c,c^2}_{\rho} = \ket{e}\bra{c^2},
\label{diracdefinitions}
\end{equation}
so we eventually obtain
\begin{equation}
F^G = 
\begin{pmatrix}
0 & 1 & \bar\omega \\
1 & 0 & \bar\omega \\
\omega & \omega & 0
\end{pmatrix}
\label{eq:FG}
\end{equation}
in the basis $\{|e\rangle, |c\rangle,|c^2\rangle\}$. The fusion of two $G$ anyons can be reproduced from \eqref{eq:FG} by multiplying the minimal ribbon operators $F^G$ together, as follows
\begin{equation}
F^G\times F^G = 
\begin{pmatrix}
2 & 1 & \bar\omega \\
1 & 2 & \bar\omega \\
\omega & \omega & 2
\end{pmatrix}
=F^A+F^B+F^G
\end{equation}
Hence, the minimal ribbon operators reproduce the non-Abelian fusion properties of $G$ anyons.

We therefore look towards the relation upon exchanging the anyons of this model, described by their $R$ matrices. With the equation for ribbon operator exchange (\ref{qdrribbondef}) and the definition of the $G$ dyon ribbon operator (\ref{gribbon}), we can find the relation
\begin{equation}
F^G_{\rho_1}F^G_{\rho_2} = R_{GG}F^G_{\rho_2}F^G_{\rho_1},
\label{eq:braid}
\end{equation}
with $R_{GG}$ being the $R$ matrix for exchanging two $G$ dyons given in \eqref{grmatrix}. 

To find $R$ from the ribbon operators, we first need to determine $F^G_{\rho_1}F^G_{\rho_2}$ and $F^G_{\rho_2}F^G_{\rho_1}$.
We can multiply two $G$ ribbon operators together
\begin{equation}
    F^G_{\rho}F^G_{\rho} = (F^{c,e}_{\rho} + \omega F^{c,c}_{\rho} + \bar\omega F^{c,c^2}_{\rho} + F^{c^2,e}_{\rho} + \bar\omega F^{c^2,c}_{\rho} + \omega F^{c^2,c^2}_{\rho})^2 
\label{gribbonsquared},
\end{equation}
which gives the result
\begin{equation}
\begin{split}
     F^G_{\rho}F^G_{\rho} &= 2F^{e,e}_{\rho} + 2F^{e,c}_{\rho} + 2F^{e,c^2}_{\rho} + F^{c,e}_{\rho} + \omega F^{c,c}_{\rho}  \\
     &\hspace{1cm} + \bar\omega F^{c,c^2}_{\rho} + F^{c^2,e}_{\rho} + \bar\omega F^{c^2,c}_{\rho} + \omega F^{c^2,c^2}_{\rho}, \\
     &= F^A_{\rho} + F^B_{\rho} + F^G_{\rho},
\end{split}
\end{equation}
as expected from the fusion rules.

From the definitions of the group ribbon operators given in (\ref{diracdefinitions}), this can be rewritten as
\begin{equation}
    F^G_{\rho_1}F^G_{\rho_2} = 
    \begin{pmatrix}
    2 & \bar\omega & \bar\omega \\
    1 & 2 & \omega \\
    1 & \omega & 2
    \end{pmatrix}
    \label{rho1rho2matrix}
\end{equation}
in the basis $\ket{e}, \ket{c}, \ket{c^2}$. To find the form of $F^G_{\rho_2}F^G_{\rho_1}$, we calculate to the full expansion of (\ref{gribbonsquared}) as
\begin{widetext}
\begin{equation}
 \begin{split}
    F^G_{\rho_1}F^G_{\rho_2} &= F^{c,e}_{\rho_1}F^{c,e}_{\rho_2} + \omega F^{c,e}_{\rho_1}F^{c,c}_{\rho_2} + \bar\omega F^{c,e}_{\rho_1}F^{c,c^2}_{\rho_2} + F^{c,e}_{\rho_1}F^{c^2,e}_{\rho_2} + \bar\omega F^{c,e}_{\rho_1}F^{c^2,c}_{\rho_2} + \omega F^{c,e}_{\rho_1}F^{c^2,c^2}_{\rho_2} \\
    &+ \omega F^{c,c}_{\rho_1}F^{c,e}_{\rho_2} + \bar\omega F^{c,c}_{\rho_1}F^{c,c}_{\rho_2} + F^{c,c}_{\rho_1}F^{c,c^2}_{\rho_2} + \omega F^{c,c}_{\rho_1}F^{c^2,e}_{\rho_2} + F^{c,c}_{\rho_1}F^{c^2,c}_{\rho_2} + \bar\omega F^{c,c}_{\rho_1}F^{c^2,c^2}_{\rho_2} \\
    &+ \bar\omega F^{c,c^2}_{\rho_1}F^{c,e}_{\rho_2} + F^{c,c^2}_{\rho_1}F^{c,c}_{\rho_2} + \omega F^{c,c^2}_{\rho_1}F^{c,c^2}_{\rho_2} + \bar\omega F^{c,c^2}_{\rho_1}F^{c^2,e}_{\rho_2} + \omega F^{c,c^2}_{\rho_1}F^{c^2,c}_{\rho_2} + F^{c,c^2}_{\rho_1}F^{c^2,c^2}_{\rho_2} \\
     &+ F^{c^2,e}_{\rho_1}F^{c,e}_{\rho_2} + \omega F^{c^2,e}_{\rho_1}F^{c,c}_{\rho_2} + \bar\omega F^{c^2,e}_{\rho_1}F^{c,c^2}_{\rho_2} + F^{c^2,e}_{\rho_1}F^{c^2,e}_{\rho_2} + \bar\omega F^{c^2,e}_{\rho_1}F^{c^2,c}_{\rho_2} + \omega F^{c^2,e}_{\rho_1}F^{c^2,c^2}_{\rho_2} \\
     &+ \bar\omega F^{c^2,c}_{\rho_1}F^{c,e}_{\rho_2} + F^{c^2,c}_{\rho_1}F^{c,c}_{\rho_2} + \omega F^{c^2,c}_{\rho_1}F^{c,c^2}_{\rho_2} + \bar\omega F^{c^2,c}_{\rho_1}F^{c^2,e}_{\rho_2} + \omega F^{c^2,c}_{\rho_1}F^{c^2,c}_{\rho_2} + F^{c^2,c}_{\rho_1}F^{c^2,c^2}_{\rho_2} \\
     &+ \omega F^{c^2,c^2}_{\rho_1}F^{c,e}_{\rho_2} + \bar\omega F^{c^2,c^2}_{\rho_1}F^{c,c}_{\rho_2} + F^{c^2,c^2}_{\rho_1}F^{c,c^2}_{\rho_2} + \omega F^{c^2,c^2}_{\rho_1}F^{c^2,e}_{\rho_2} + F^{c^2,c^2}_{\rho_1}F^{c^2,c}_{\rho_2} + \bar\omega F^{c^2,c^2}_{\rho_1}F^{c^2,c^2}_{\rho_2}.
\end{split}
\end{equation}
Using (\ref{qdrribbondef}) we can exchange these ribbons to get
\begin{equation}
 \begin{split}
    F^G_{\rho_2}F^G_{\rho_1} & = \omega F^{c,e}_{\rho_2}F^{c,e}_{\rho_1} + \bar\omega F^{c,e}_{\rho_2}F^{c,c}_{\rho_1} + F^{c,e}_{\rho_2}F^{c,c^2}_{\rho_1} + \bar\omega F^{c,e}_{\rho_2}F^{c^2,e}_{\rho_1} + \omega F^{c,e}_{\rho_2}F^{c^2,c}_{\rho_1} +  F^{c,e}_{\rho_2}F^{c^2,c^2}_{\rho_1} \\
    &+ \bar\omega F^{c,c}_{\rho_2}F^{c,e}_{\rho_1} + F^{c,c}_{\rho_2}F^{c,c}_{\rho_1} + \omega F^{c,c}_{\rho_2}F^{c,c^2}_{\rho_1} +  F^{c,c}_{\rho_2}F^{c^2,e}_{\rho_1} + \bar\omega F^{c,c}_{\rho_2}F^{c^2,c}_{\rho_1} + \omega F^{c,c}_{\rho_2}F^{c^2,c^2}_{\rho_1} \\
    &+  F^{c,c^2}_{\rho_2}F^{c,e}_{\rho_1} + \omega F^{c,c^2}_{\rho_2}F^{c,c}_{\rho_1} + \bar\omega F^{c,c^2}_{\rho_2}F^{c,c^2}_{\rho_1} + \omega  F^{c,c^2}_{\rho_2}F^{c^2,e}_{\rho_1} +  F^{c,c^2}_{\rho_2}F^{c^2,c}_{\rho_1} + \bar\omega F^{c,c^2}_{\rho_2}F^{c^2,c^2}_{\rho_1} \\
     &+ \bar\omega F^{c^2,e}_{\rho_2}F^{c,e}_{\rho_1} +  F^{c^2,e}_{\rho_2}F^{c,c}_{\rho_1} + \omega F^{c^2,e}_{\rho_2}F^{c,c^2}_{\rho_1} + \omega F^{c^2,e}_{\rho_2}F^{c^2,e}_{\rho_1} + F^{c^2,e}_{\rho_2}F^{c^2,c}_{\rho_1} + \bar\omega F^{c^2,e}_{\rho_2}F^{c^2,c^2}_{\rho_1} \\
     &+ \omega F^{c^2,c}_{\rho_2}F^{c,e}_{\rho_1} + \bar\omega F^{c^2,c}_{\rho_2}F^{c,c}_{\rho_1} + F^{c^2,c}_{\rho_2}F^{c,c^2}_{\rho_1} +  F^{c^2,c}_{\rho_2}F^{c^2,e}_{\rho_1} + \bar\omega F^{c^2,c}_{\rho_2}F^{c^2,c}_{\rho_1} + \omega F^{c^2,c}_{\rho_2}F^{c^2,c^2}_{\rho_1} \\
     &+ F^{c^2,c^2}_{\rho_2}F^{c,e}_{\rho_1} + \omega F^{c^2,c^2}_{\rho_2}F^{c,c}_{\rho_1} + \bar\omega F^{c^2,c^2}_{\rho_2}F^{c,c^2}_{\rho_1} + \bar\omega F^{c^2,c^2}_{\rho_2}F^{c^2,e}_{\rho_1} + \omega F^{c^2,c^2}_{\rho_2}F^{c^2,c}_{\rho_1} +  F^{c^2,c^2}_{\rho_2}F^{c^2,c^2}_{\rho_1}, 
\end{split}
\end{equation}
\end{widetext}
with each term gaining a phase of either $\omega$ or $\bar\omega$ when compared to the above. We choose now the two paths to be on the top of each other, $\rho_1=\rho_2=\rho$ and employ \eqref{diracdefinitions} and \eqref{eq:con} to find
\begin{equation}
F^G_{\rho_2}F^G_{\rho_1} = 
\begin{pmatrix}
2\bar\omega & \omega & 1 \\
\omega & 2\bar\omega & 1 \\
\bar \omega & \bar\omega & 2\bar\omega
\end{pmatrix} = \bar \omega (F^A +F^B) + \omega F^G.
\end{equation}
Comparing this to (\ref{rho1rho2matrix}), it is seen that the diagonal terms have gained a phase $\bar\omega$ and the off-diagonal have gained a phase $\omega$. Noting that the diagonal terms come from the fusion of $G \times G \to A + B$ and the off-diagonal are from $G \times G \to G$, it is evident that the $A$ and $B$ fusion channels have gained a phase $\bar\omega$ and the $G$ has gained a phase $\omega$. This is the expected result from the $R$ matrix in the anyonic basis (\ref{grmatrix}).

\setcounter{equation}{0}
\setcounter{figure}{0}
\setcounter{table}{0}

\section{Appendix}
% \textcolor{red}{\textbf{\textit{Non-unitarity:}}  [[All information related to this section was moved to main text, but can be moved here later]]}

% % It is known that it is impossible to implement a non-unitary operation using linear optics efficiently.
% This poses a challenge since the operators of interest are non-unitary.
% Implementing such operations leaves us with a choice between the accuracy of the operation, and the efficiency of the operation.
% This task can be mapped to the problem of state discrimination where the task is to sort non-orthogonal quantum states, where each column of the operation corresponds to a state that needs to be discriminated.
% Since in this work we are simply simulating these Hermitian operators, we choose the accuracy of the operations while adding loss to the overall implementation.
% Hence, this problem can be simply mapped to high-dimensional unambiguous state discrimination, where auxiliary modes are used to accurately sort overlapping quantum states. 
% This problem can be further simplified if the overlap between the optical modes is symmetric, i.e. $| \bra{\psi_i}\psi_{j\neq i}\rangle |^2=\alpha$.
% Thankfully, the three Hermitian operators $\mathbf{T} = \{$ \Ha, \Hb , \Hc$ \}$ are such that they have symmetric overlaps between different columns, i.e. $|T_{rj} \cdot {T_{r{i\neq j}}}^*|^2 = \alpha$ where $T_{rc}$ corresponds to the $r^{th}$ row and $c^{th}$ column of the given $\mathbf{T}$ matrix. 
% This allows us to use a single auxiliary mode to perform these operations as recently shown in \cite{goel2022simultaneously}.
\subsection{Experimental Apparatus}\label{append:apparatus}

As shown in Fig.~\ref{fig:apparatus}, our apparatus consists of a coherent light source at $810$ nm launched using a single mode fiber (Thorlabs-780HP) collimated using a set of lenses with an effective focal length of $L_1 = 58.67$ mm. The beam is reflected off a spatial light modulator (Hamamatsu LCOS-X10468, $\text{SLM}_1$) and coupled into a multi-mode GRIN fibre (Thorlabs-M116L02)
with a set of lenses with an effective focal length of $L_2 = 22$ mm. The light from the multi-mode fibre is then incident on another spatial light modulator ($\text{SLM}_2$) with the help of another set of lenses with an effective focal length of $L_3 = 22$mm. Finally, the modulated light is measured on a CMOS Camera (XIMEA-xiC USB3.1) after passing through a set of lenses with an effective focal length of $L_4=33.33$ mm. %The SLMs used in this experiment are commercially available Hamamatsu LCOS-X10468. 

\subsection{Inverse design of operations}\label{append:tdc}

We encode the desired operations in an optical circuit that is built with a cascade of optical mode mixers separated by phase layers as described in \cite{goel2022inverse}. For a given set of mode mixers $U_j$ and phase layers $P_j$, a given operation $\mathbf{T}$  is encoded as
\be
    \mathbf{T} = \prod_{j=1}^{2}U_j P_j.
    \label{eq_tdc}
\ee

Here the mode mixers $U_1$ and $U_2$ correspond to an MMF and a $2f$ lenses system respectively. The phase layers $P_j$ correspond to the SLMs.
In order to encode these operations to the given optical circuit, we perform an inverse design algorithm called wavefront-matching such that a set of input optical modes are transformed according to the given transformation to a set of output modes. The inverse design of operations requires measurement of the transmission matrix (TM) of the MMF, which is done by displaying a set of random patterns on both $\text{SLM}_1$ and $\text{SLM}_2$ and subsequently measuring the intensity of the output speckle patter on the camera. These random measurements serve as a data-set to train a multi-plane neural network (MPNN) which reveals the TM of the MMF~\cite{goel2023TMML}.

The output modes for the implemented operations are chosen such that the measurement outcomes are spatially separated and can be measured using a camera. The position of spatially-separated measurement outcomes on the camera can be arbitrarily selected, however, might result in different performances. We thus implement multiple realisations of the same operation with different positions of measurement outcomes on the camera.

\subsection{Experimentally implemented processes}\label{append:quantum_processes}

Experimentally, when attempting to implement the desired operations, there are additional incoherent effects which must be characterised by determining the entire quantum process, which describes the map between density operator on the input space to those on the output. The implemented processes take the general form $\mathcal{\widetilde{T}}(\rho)= \sum_n^R \mathbf{\widetilde{T}}_n \rho \mathbf{\widetilde{T}}_n^\dagger$. This is the Kraus representation of the process and it is completely positive (CP) by construction. We do not require that our operations are trace preserving (TP) because loss may, and indeed must, occur in the system. In this case the process must be non-trace-increasing, and the Kraus operators, $\{\mathbf{\widetilde{T}}_n\}_n$, of the process must obey $\sum_n \mathbf{\widetilde{T}}^\dagger _n\mathbf{\widetilde{T}}_n\leq\mathbb{I}$, ie. no amplification can occur. Loss could be accounted for by introducing an auxiliary output mode and an additional Kraus operator, and imposing equality $\sum_n \mathbf{\widetilde{T}}^\dagger _n\mathbf{\widetilde{T}}_n=\mathbb{I}$ to arrive at a TP process. 
 The ideal operations correspond to the processes $\mathcal{T}(\rho)= \mathbf{T} \rho \mathbf{T}^\dagger$, which are, again, not TP. The non-trace-increasing condition is saturated for operations $\mathbf{T}/\alpha$ when $\alpha$ is chosen to be the largest singular value of $\mathbf{T}$. In this case the average success probability of the operations, defined as $\tr [ (\mathbf{T}/\alpha) (\mathbb{I}/3) (\mathbf{T}/\alpha)^\dagger]$, is given by $1/2, \,3/8,\, 3/4$ for \Ha, \Hb , \Hc , respectively.
We perform quantum process tomography (QPT) to characterise the implemented processes $\mathcal{\widetilde{T}}$ and quantify their fidelity in relation to the ideal operations $\mathcal{T}$. 
As an alternative to the Kraus representation, it is convenient to represent these processes via their Choi states, $\rho_{\mathcal{\widetilde{T}}}= \mathcal{\widetilde{T}} \otimes \mathbb{1} (\rho^+)$, with $\rho^+$ the maximally entangled state.
Rather than work with the non-TP Choi state, we normalise to unit trace, thus providing the appropriate object for assessing process fidelities whilst neglecting global losses~\cite{Bongioanni2010}.
The purity of the implemented transformation $\mathcal{\widetilde{T}}$ is calculated as $\mathcal{P} = Tr(\rho_{\widetilde{\mathcal{T}}}^2)$~, while its fidelity with respect to the ideal transformation $\mathbf{T}$ is calculated as,
\be
\mathcal{F}(\rho^{}_{\mathcal{T}},\rho_{\widetilde{\mathcal{T}}}) = \left[\text{Tr} \left(\sqrt{ \sqrt{\rho_{\mathcal{T}}} \rho_{\widetilde{\mathcal{T}}} \sqrt{\rho_{\mathcal{T}}}} \right)\right]^2,
\label{Eq:F}
\ee

\subsection{Quantum process tomography (QPT) of implemented processes}\label{append:p_tomo}

 In order to tomograph the implemented operations, we prepare and measure each vector from an informationally complete set of mutually unbiased bases(MUB)~\cite{Giovannini2013}. The $j^\text{th}$ vector in the $\mu^\text{th}$ MUB in three dimensions is given by~\cite{Wootters1989} 
 
\be
    \ket{M^\mu_{j}}=\frac{1}{\sqrt{3}}\sum^{2}_{i=0}\omega^{ji+\mu i^2}\ket{i}
    \label{eqn:MUB}
\ee

 where $\ket{i}$ corresponds to $i^\text{th}$ mode in the computational basis, and $\omega=e^{i \pi / 3}$ is the cube root of unity.
 Experimentally, we prepare each of the input states, $\{ \hat \tau^\mu_j =\ket{M^\mu_{j}}\bra{M^\mu_{j}} \}_{\mu j}$, as superpositions of macro-pixels using $\text{SLM}_1$ in all MUBs as shown in Fig.~\ref{fig:apparatus}(b). For each of the prepared input states, the output of the optical circuit is projected on all MUB vectors,  $\{ \hat \Pi^\mu_j =\ket{M^\mu_{j}}\bra{M^\mu_{j}} \}_{\mu j}$, of the modes corresponding to the three selected output modes using $\text{SLM}_2$.  These measurements are, up to a constant intensity, $N$, proportional to the expectation values,
 
 \be
    I^{\mu \nu}_{i j} = N \tr[\hat \Pi^\mu_i \chi(\hat \tau^\nu_j)]\, ,
    \label{eqn:PTmeas}
\ee

with $\chi$ the quantum process being implemented by the optical circuit acting on input state $\hat \tau^\mu_j$ being projected onto $\hat \Pi^\mu_j$, which are sufficient for QPT. In order to recover the physical quantum process, $\chi$, from this data, we impose positivity through the positivity of the Choi state ($\rho_\chi:= \chi \otimes \mathbb{1} (\rho^+)$, with $\rho^+$ the maximally entangled state) representation of the process allowing QPT to be performed via the semi-definite program (SDP),
 
 \be
 \begin{split}
 \label{eqs:QSTSDP}
\min_{\rho_\chi, N} \quad& |I^{\mu \nu}_{i j}-N \tr[\hat \Pi^\mu_i \otimes \hat (\tau^\nu_j)^T \rho_\chi]|^2 \\
\textrm{s.t.} \quad & \rho_\chi\geq 0 \, , \, \tr[\rho_\chi]=1.
\end{split}
\ee

 In this manner, the recovered Choi state of the process is not TP, and normalised to unit trace suitable for assessing process fidelities.

\subsection{Systematic error analysis due to misalignments}\label{append:error-analysis}

 In practice, the states prepared and measurements performed deviate from the ideal cases, which can lead to erroneous process reconstruction. The extent, and particularly the distribution, of the infidelity of state preparation and measurement can by characterised experimentally by sampling from independent prepare and measure experiments, and fitting these data to a model capturing sources of systematic error in the experimental devices~\cite{goel2022inverse}. This model, and the distributions of different types of systematic error, allow to simulate data sampling from these imperfections, perform the QPT procedure to obtain a Monte-Carlo sample of these effects, and arrive at realistic bounds on the systematic errors present on our process fidelities. These errors upto three standard deviations are presented in Table.~\ref{tab:res} .

\setcounter{equation}{0}
\setcounter{figure}{0}
\setcounter{table}{0}

\section{Appendix}\label{append:block-encoding}
\subsection{An explicit qubit block encoding based on the SVD}
We wish to construct a unitary block encoding \cite{low2017optimal,gilyen2019quantum,camps2022explicit,low2019hamiltonian,camps2022fable} of the ribbon operator $F_{\rho_0}^G$ in the top-left block of a unitary matrix:
\begin{equation}
    U_F = 
    \left(
        \begin{array}{cc}
            F_{\rho_0}^G/\alpha & \star   \cr
            \star & \star  \cr
        \end{array}
    \right),
\end{equation}
where $\alpha$ is a scaling factor such that in terms of the operator norm $||F_{\rho_0}^G/\alpha||\leq 1$, and $\star$ denotes arbitrary matrix elements that only have that constraint that that the total matrix  $U_F$ is unitary.

We now give an explicit unitary block encoding of $F_{\rho_0}^G$ based on the  singular-value decomposition (SVD) of $F_{\rho_0}^G$. The SVD is of the form $ F_{\rho_0}^G/\alpha = U_1 D U_2$ where $D$ is a diagonal matrix and where both $U_1$ and $U_2$ are two unitary matrices of the same dimension as $F_{\rho_0}^G$. If we have $||F_{\rho_0}^G/\alpha||\leq 1$ then the following matrix is guaranteed to be unitary \cite{horn2012matrix,camps2022explicit}:
\begin{equation}
  U_F=  \begin{pmatrix}
          F_{\rho_0}^G/\alpha & U_1 \sqrt{\mathds{1} - D^2} U_2   \\
     U_1 \sqrt{\mathds{1} - D^2} U_2 & - F_{\rho_0}^G/\alpha 
    \end{pmatrix}
\end{equation}

As $F_{\rho_0}^G$ is low dimensional, it is straightforward to calculate $U_F$ numerically. This gives:
\begin{widetext}
\begin{equation}\label{eqn:unit-embed}
    U_F =  \bordermatrix{~ 
    & \bra{000} & \bra{100} & \bra{010} & \bra{110} & \bra{001} & \bra{101} & \bra{011} & \bra{111} \cr
    \ket{000} &      0 & \frac{1}{2} & \frac{1}{2}\bw & 0 & -\frac{1}{\sqrt{3}} & \frac{1}{2 \sqrt{3}} & \frac{ \bw}{2 \sqrt{3}} & 0 \cr
    \ket{100} &     \frac{1}{2} & 0 & \frac{1}{2} \bw & 0 & \frac{1}{2 \sqrt{3}} & -\frac{1}{\sqrt{3}} & \frac{ \bw}{2 \sqrt{3}} & 0 \cr
    \ket{010} &     \frac{1}{2} \w & \frac{1}{2} \w & 0 & 0 & \frac{\w}{2 \sqrt{3}} & \frac{\w}{2 \sqrt{3}} & -\frac{1}{\sqrt{3}} & 0 \cr
    \ket{110} &      0 & 0 & 0 & 0 & 0 & 0 & 0 & 1 \cr
    \ket{001} &     -\frac{1}{\sqrt{3}} & \frac{1}{2 \sqrt{3}} & \frac{ \bw}{2 \sqrt{3}} & 0 & 0 & -\frac{1}{2} & \frac{- \bw}{2} & 0 \cr
    \ket{101} &     \frac{1}{2 \sqrt{3}} & -\frac{1}{\sqrt{3}} & \frac{ \bw}{2 \sqrt{3}} & 0 & -\frac{1}{2} & 0 & \frac{- \bw}{2} & 0 \cr
    \ket{011} &     \frac{\w}{2 \sqrt{3}} & \frac{\w}{2 \sqrt{3}} & -\frac{1}{\sqrt{3}} & 0 & -\frac{1}{2} \w & -\frac{1}{2} \w & 0 & 0 \cr
    \ket{111} & 0 & 0 & 0 & 1 & 0 & 0 & 0 & 0 \cr
         }
\end{equation}
\end{widetext}
where $\alpha = 2$, and where we pad $F_{\rho_0}^G$ with zeroes.

\subsection{Success probability of qubit block encoding}
The following circuit then allows us apply $F_{\rho_0}^G$ to the encoded qutrit state $\ket{\phi}$, using $U_F$ within the 3 qubit system:
\begin{equation}
    \Qcircuit @C=1em @R=1em {
        \lstick{\ket{0}} & \multigate{1}{U_F} & \meter  \qw \\
        \lstick{\ket{\phi}} & \ghost{U_F} & \rstick{\propto F_{\rho_0}^G \ket{\phi}} \qw
    }
\end{equation}
if the measurement of the first qubit returns 0.
% or in dirac notation:
% \begin{equation}
%     \alpha ( \bra{0} \x \I) U_A (\ket{0} \x \I ) = A.
% \end{equation}
Success probably is therefore dependent on the input state, as
% \begin{equation}
%     p(0) = \frac{1}{\alpha^2} || F_{\rho_0}^G \ket{\phi}||^2 = \frac{1}{\alpha^2} \bra{\phi} (F_{\rho_0}^{G} )^{\dag} F_{\rho_0}^G \ket{\phi}.
% \end{equation}
% Relaxing the input to mixed states 
we have $p(0) = \frac{1}{\alpha^2} \tr[(F_{\rho_0}^{G} )^{\dag} F_{\rho_0}^G \rho ] $ and therefore we have the following bounds on the success probability of the block-encoding
\begin{equation}
     \min_k ( \sigma_k^2 ) \leq p(0) \leq  \max_k ( \sigma_k^2 ),
\end{equation}
where $\sigma_k$ are the singular values of $F_{\rho_0}^{G}/\alpha$. Given a minimal required value of $\alpha = 2$ for positivity of $F_{\rho_0}^{G}/\alpha$, we have $\{\sigma_k\} = \{1,1/2,1/2\}$ and therefore the success probability is bounded tightly as
\begin{equation}
    \frac{1}{4} \leq p(0) \leq 1.
\end{equation}
We can find the states that saturate these bounds through the eigen-decomposition of $(F_{\rho_0}^{G} )^{\dag} F_{\rho_0}^G$, which is clearly closely related to the above singular values. We find $e_1=1$, $e_2=1/4$, $e_3=1/4$, with normalized eigenvectors of 
\begin{equation}
    \ket{e_1} = \frac{1}{\sqrt{3}} \begin{pmatrix} \bw \\ \bw \\ 1 \end{pmatrix}, \ \ket{e_2} = \frac{1}{\sqrt{2}} \begin{pmatrix} -\bw \\ 0 \\ 1 \end{pmatrix}, \ \ket{e_3} = \frac{1}{\sqrt{2}} \begin{pmatrix} -1 \\ 1 \\ 0 \end{pmatrix},
\end{equation}
where $\ket{e_1}$ saturates the upper bound, and $\ket{e_2}$ and $\ket{e_3}$ the lower bound. Appending five zeroes to these vectors gives the required three qubit states, $\ket{\psi_i}$, in the same basis as $U_F$ in Equation (\ref{eqn:unit-embed}).

\subsection{Fidelity of IBM device implementation}
We now consider how well a current IBM device might implement the circuits applying $U_F$ on the three eigenstates. Using the standard native gates of the IBMQ device, here \textsf{ibmq\_belem}, we transpile circuits to prepare the eigenstates, and apply the unitary $U_F$. Using the qiskit complier we find $U_F$ requires a circuit depth of 74 operations with 20 CNOTs.

Through IBMQ we can access a simplified model version of the noise that has been characterised on a device. For \textsf{ibmq\_belem} we download the noise model associated with its native gates, and simulate our circuits with and without noise. We can then compare the density matrices of the idealised case with the noisy versions. We denote the ideal unitary channel $\mathcal{U}_F(\cdot) := U_F \cdot U_F^\dagger$, and denote the noisy version, $\tilde{\mathcal{U}}_F$. For the embedded eigenstates,  $\psi_i := |\psi_i\rangle \langle \psi_i |$, we denote the noisy versions, $\rho_i$. The results of this are shown in Table \ref{tab:result-qiskit}. We see that the fidelities between the idealized implementation and simulated noisy implementation are strictly below $81\%$, suggesting that current quantum devices cannot achieve the high levels of fidelity of the photonic simulation.

\begin{table}[!h]
    \centering
    \begin{tabular}{|c||c|c||c|c|} 
        \hline
         State & \begin{tabular}{@{}c@{}}Initial \\ Purity\end{tabular}  & Initial Fidelity &  \begin{tabular}{@{}c@{}}Final \\ Purity\end{tabular} &  Final Fidelity  \\
         $ \psi_i $ &   $\mathcal{P}(\rho_i)$ &   $\mathcal{F}(\psi_i,\rho_i)$ &   $\mathcal{P}(\tilde{\mathcal{U}_F}(\rho_i))$ &   $\mathcal{F}(\mathcal{U}_F,\tilde{\mathcal{U}}_F)$  \\[1mm]
         \hline
 i=1  & $92.9\%$ & $96.4\%$ & $66.0\%$ & $80.9\%$\\ [1mm] 
 i=2 & $92.7\%$ & $96.3\%$ & $62.3\%$ & $78.4\%$ \\ [1mm] 
  i=3   & $91.5\%$ & $95.6\%$ & $61.6\%$ & $77.8\%$ \\ [1mm] 
         \hline
    \end{tabular}
    \caption{Simulated results comparing noisy quantum circuits of the 3 qubit unitary block encoding of $F_{\rho_0}^G$ to idealised noise free case. We use the noise model of the gateset associated to the device \textsf{ibmq\_belem}.}
     \label{tab:result-qiskit}
\end{table}